\documentstyle[preprint,seceq,epsf,rotate]{ptptex}

\def \ie {{\it i.e.}}
 
\newcommand {\both}[1]	{\relax\ifmmode#1\else$#1$\fi\relax}

\newcommand {\baresq}	{\bar{e}^2}
\newcommand {\bargsq}	{\bar{g}^2}
\newcommand {\barssq}	{\bar{s}^2}

\newcommand {\bargzsq}	{\bar{g}_Z^2}
\newcommand {\bargwsq}	{\bar{g}_W^2}

\newcommand {\hatesq}	{\hat{e}^2}
\newcommand {\hatgsq}	{\hat{g}^2}
\newcommand {\hatssq}	{\hat{s}^2}
\newcommand {\hatcsq}	{\hat{c}^2}
\newcommand {\hatgzsq}	{\hat{g}_Z^2}

\newcommand {\eeww} {$e^+e^- \to W^+W^-$\/}

\newcommand{\tr}{{\rm Tr}\,}

\newcommand{\cm}{\raisebox{-0.2cm}{\bf X}}
\newcommand{\cmo}{\raisebox{-0.2cm}{\bf O}}

\newcommand{\dcov}{{D}}

\newcommand{\id}{{\bf 1}}

\preprintnumber{
\hspace*{-1cm}KEK-TH-465\\\hspace*{-1cm}KEK Preprint 95-193}

\title{%
A comparison of the non-standard effects of effective theories with 
linearly realized and strongly interacting Higgs sectors\footnote{Talk 
presented at the 1995 Yukawa International Seminar, Kyoto, Japan,
21-25 August, 1995.}
}

\author{%
Robert F. Szalapski
}

\inst{%
Theory Group, KEK, 1-1 Oho, Tsukuba, Ibaraki 305, Japan\\
E-mail address: robs@theory.kek.jp
}

\recdate{%
\today
}

\abst{%
Effective theories provide a powerful tool for testing the Standard Model
and for searching for the effects of new physics in a model-independent manner.
In general one assumes that the effects of new physics characterized by a 
high-energy scale may be observed at low-energies only through quantum 
corrections.  These quantum corrections may be described by effective operators
which are constructed from the fields of the low-energy theory; the coefficients
of these operators may be measured via experiment, but a theoretical 
determination is only possible once the complete theory is known.  Because the
existence or non-existence of a light physical Higgs boson has not yet been 
established the inclusion or exclusion of the Higgs field in the construction 
of the effective Lagrangian is open to debate.  For this reason we discuss both
the linearly realized effective Lagrangian, which includes a light physical 
Higgs boson, and the chiral Lagrangian, where the Higgs boson has been removed
by a formal integration.  In each scenario the contributions to oblique 
parameters and to W-pair-production amplitudes are discussed.  Where possible
relationships between the two effective Lagrangians are given.
}

\begin{document}

\maketitle


\section{The linear representation and energy-dimension-six 
operators}\label{sec-linear}

If the scale of new physics, $\Lambda$, is large compared to the vacuum 
expectation value (vev) of the the Higgs field, $v = 246.22$GeV, then the 
effective Lagrangian may be expressed as the SM Lagrangian plus terms with 
energy dimension greater than four suppressed by inverse powers of $\Lambda$, 
\ie
\begin{equation}\label{fulllagrangian}
{\cal L}_{eff} = {\cal L}_{\rm SM} 
+  \sum_{n \geq 5} \sum_i \frac{f_i^{(n)} \, 
{\cal O}_i^{(n)}}{\Lambda^{n-4}} \;.
\end{equation}
The energy dimension of each operator is denoted by $n$, and the index $i$ sums 
over all operators of the given energy dimension.  The coefficients $f_i^{(n)}$
are free parameters.

In this section we assume that the low-energy theory, \ie the SM, contains a 
light physical scalar Higgs particle which is the remnant of a complex 
Higgs-doublet field.  We assume that the couplings of the new physics to 
fermions are suppressed, and we only consider operators which conserve CP.  Upon 
restricting the analysis to operators not exceeding energy-dimension six we find 
that twelve operators form a basis set; all are dimension-six and separately 
conserve C and P\cite{HISZnew}.  
They are summarized in Table~\ref{table-linear}.  
For convenience when defining the normalizations of the individual operators 
we use the `hatted' field strength tensors defined according to
\begin{equation}
\Big[ D_\mu, D_\nu \Big] = \hat{W}_{\mu\nu} + \hat{B}_{\mu\nu}
= i g T^a W^a_{\mu\nu} + i g^\prime Y B_{\mu\nu}\;. \label{comu}
\end{equation}
Combining the twelve operators of Table~\ref{table-linear} with 
Eqn.~(\ref{fulllagrangian}) completes the construction of the effective 
Lagrangian in the linear representation.  

Also in Table~\ref{table-linear} we indicate those vertices to which each 
operator contributes with an `{\cal X}' in the appropriate box.  
First, observe that four of the operators, ${\cal O}_{DW}$,
${\cal O}_{DB}$, ${\cal O}_{BW}$ and ${\cal O}_{\Phi,1}$, contribute to 
gauge-boson two-point-functions at the tree level.  For this reason their 
respective coefficients, $f_{DW}$, $f_{DB}$, $f_{BW}$ and $f_{\Phi,1}$, are
strongly constrained by LEP/SLC and low-energy data\cite{HMS}, and these 
constraints will be improved by data at higher energy lepton colliders.   (This 
will be discussed in greater detail in Sec.~\ref{sec-eeff}.)  These four
operators will contribute to the process \eeww\/ through corrections to the 
charge form-factors, $\baresq(q^2)$, $\barssq(q^2)$, $\bargzsq(q^2)$ and  
$\bargsq(q^2)$, and through the W-boson wave-function-renormalization factor,
$Z_W^{1/2}$.

The operators ${\cal O}_{DW}$ and ${\cal O}_{BW}$ also make a direct 
contribution to $WW\gamma$
and $WWZ$ vertices.  Three additional operators contribute as well.  They are 
${\cal O}_{WWW}$, ${\cal O}_W$ and ${\cal O}_B$; their respective 
coefficients are $f_{WWW}$, $f_W$ and $f_B$.  The non-zero 
direct contributions may be summarized by
\begin{subequations}
\label{linear-direct}
\begin{eqnarray}
\label{deltag1gammadirect}
\Delta g^\gamma_{1,{\rm direct}} (q^2) & = & 2\hatgsq 
  \frac{q^2 + 2 \hat{m}_W^2}{\Lambda^2}f_{DW} \;, \\
\label{deltag1zdirect}
\Delta g^Z_{1,{\rm direct}} (q^2) & = & 2\hatgsq 
  \frac{q^2 + 2 \hat{m}_W^2}{\Lambda^2}f_{DW} 
  + \frac{1}{2}\frac{\hat{m}_Z^2}{\Lambda^2}f_W \;, \\
\label{deltakappagammadir}
\Delta \kappa_{\gamma,{\rm direct}}  (q^2) & = & 2\hatgsq 
  \frac{q^2 + 2 \hat{m}_W^2}{\Lambda^2}f_{DW} 
  + \frac{1}{2}\frac{\hat{m}_W^2}{\Lambda^2}
  \Big(f_W -2 f_{BW}  + f_B\Big) \;, \\
\label{deltakappazdir}
\Delta \kappa_{Z,{\rm direct}}  (q^2) & = & 2\hatgsq 
  \frac{q^2 + 2 \hat{m}_W^2}{\Lambda^2}f_{DW} 
  + \frac{1}{2}\frac{\hat{m}_Z^2}{\Lambda^2}
  \Big(\hatcsq f_W +2\hatssq f_{BW} - \hatssq f_B\Big)\;, \\
\nonumber
\Delta \lambda_{\gamma,{\rm direct}} (q^2)  
 & = & \Delta \lambda_{Z,{\rm direct}} (q^2) \\  
\label{deltalambdadirect}
& = & 
  \frac{3}{2}\hatgsq \Big(-4 f_{DW} + f_{WWW} \Big)
  \frac{\hat{m}_W^2}{\Lambda^2} \;.
\end{eqnarray}
\end{subequations}
Electromagnetic gauge invariance requires that $\Delta g^\gamma_{1}$ should 
vanish for on-shell photons, in apparent contradiction with 
(\ref{deltag1gammadirect}); the apparent contradiction will be resolved when all 
effects are included.

Naively one would expect contributions to (\ref{linear-direct}) and to the 
gauge-boson two-point-functions from 
${\cal O}_{WW}$ and ${\cal O}_{BB}$.  However, their contributions may 
be completely absorbed by a redefinition of SM fields and gauge couplings,
leading to a null contribution.  For this reason an 
`{\cal O}' is used for these operators in Table~\ref{table-linear}.
Additionally ${\cal O}_{\Phi,4}$ contributes to the W- and Z-mass terms, 
while ${\cal O}_{\Phi,1}$ contributes to the Z-mass term only.  Hence
${\cal O}_{\Phi,1}$ violates the custodial symmetry, and the $T$ parameter is 
explicitly dependent upon $f_{\Phi,1}$.  On the other hand, the contributions 
from ${\cal O}_{\Phi,4}$ exactly cancel in the calculation of $T$, hence it 
does not contribute.
\begin{table}[htbp]
\begin{center}
\begin{tabular}{|l||l|l|l|l|l|l|l|l|l|l|l|}
\hline
\makebox[1cm]{}\raisebox{5mm}{${\cal O}^{(6)}_i$} &
\rotate[l]{\makebox[1.8cm][c]{WW}} & \rotate[l]{\makebox[1.8cm][c]{ZZ}} & 
\rotate[l]{\makebox[1.8cm][c]{AZ}} & \rotate[l]{\makebox[1.8cm][c]{AA}} & 
\rotate[l]{\makebox[1.8cm][c]{WWZ}} & \rotate[l]{\makebox[1.8cm][c]{WWA}} & 
\rotate[l]{\makebox[1.8cm][c]{WWWW}} & \rotate[l]{\makebox[1.8cm][c]{WWZZ}} & 
\rotate[l]{\makebox[1.8cm][c]{WWZA}} & \rotate[l]{\makebox[1.8cm][c]{WWAA}} & 
\rotate[l]{\makebox[1.8cm][c]{ZZZZ}} \\ \hline \hline
\raisebox{-0.2cm}{${\cal O}_{DW}  =  
{\rm Tr} \bigg( \Big[ D_\mu,\hat{W}_{\nu\rho} \Big] \, 
\Big[ D^\mu,\hat{W}^{\nu\rho} \Big] \bigg)$  } 
&\cm &\cm &\cm &\cm &\cm &\cm &\cm 
&\cm &\cm &\cm & \makebox[0.4cm]{} \\[0.4cm] \hline
\raisebox{-0.2cm}{${\cal O}_{DB}  =  -\frac{g'^2}{2}\Big(\partial_\mu
B_{\nu\rho}\Big)\Big(\partial^\mu B^{\nu\rho}\Big)$ } &
& \cm & \cm & \cm &&&&&&& \\[0.4cm] \hline
\raisebox{-0.2cm}{${\cal O}_{BW} = \Phi^ \dagger \hat{B}_{\mu\nu} 
\hat{W}^{\mu\nu} \Phi$ } &
& \cm & \cm & \cm & \cm & \cm &&&&& \\[0.4cm] \hline
\raisebox{-0.2cm}{${\cal O}_{\Phi,1} = 
\bigg[ \Big(D_\mu\Phi\Big)^\dagger \Phi\bigg] \; 
\bigg[ \Phi^\dagger \Big(D^\mu \Phi\Big)\bigg] $} &
& \cm &&&&&&&&& \\[0.4cm] \hline
\raisebox{-0.2cm}{${\cal O}_{WWW} = 
\tr \Big( \hat{W}_{\mu\nu} \hat{W}^{\nu\rho} \hat{W}_\rho\,^\mu \Big) $} &
&&&& \cm & \cm & \cm & \cm & \cm & \cm & \\[0.4cm] \hline
\raisebox{-0.2cm}{${\cal O}_{WW}  =  \Phi^\dagger \hat{W}_{\mu\nu} 
\hat{W}^{\mu\nu} \Phi $} &
\cmo & \cmo & \cmo & \cmo & \cmo & \cmo & \cmo & \cmo & \cmo & \cmo 
& \\[0.4cm] \hline
\raisebox{-0.2cm}{${\cal O}_{BB} = 
\Phi^\dagger \hat{B}_{\mu\nu} \hat{B}^{\mu\nu} \Phi  $ } &
\cmo & \cmo & \cmo & \cmo & \cmo & \cmo & \cmo & \cmo 
& \cmo & \cmo & \\[0.4cm] \hline
\raisebox{-0.2cm}{$ {\cal O}_W = \Big( D_\mu\Phi \Big)^\dagger \hat{W}^{\mu\nu} 
\Big( D_\nu\Phi \Big)$ } &
&&&& \cm & \cm & \cm & \cm & \cm && \\[0.4cm] \hline
\raisebox{-0.2cm}{${\cal O}_B = \Big( D_\mu\Phi \Big)^\dagger \hat{B}^{\mu\nu}
\Big( D_\nu\Phi \Big) $ } &
&&&& \cm & \cm &&&&& \\[0.4cm] \hline
\raisebox{-0.2cm}{${\cal O}_{\Phi,2} = 
\frac{1}{2} \partial_\mu \Big( \Phi^\dagger \Phi \Big)
\partial^\mu \Big( \Phi^\dagger \Phi \Big)   $ } &
&&&&&&&&&& \\[0.4cm] \hline
\raisebox{-0.2cm}{${\cal O}_{\Phi,3}  = 
 \frac{1}{3} \Big( \Phi^\dagger \Phi \Big)^3  $ } &
&&&&&&&&&& \\[0.4cm] \hline
\raisebox{-0.2cm}{$ {\cal O}_{\Phi,4}  =  \Big( \Phi^\dagger \Phi \Big) 
\bigg[\Big( D_\mu \Phi \Big)^\dagger \Big( D^\mu \Phi \Big)\bigg]  $ } &
 \cmo & \cmo &&&&&&&&& \\[0.4cm]
\hline
\end{tabular}
\vspace*{0.2cm}
\caption{Energy-dimension-six operators in the linear representation of the 
Higgs mechanism.  
The contribution of an operator to a particular vertex is denoted 
by an \bf `X' \rm.  
In some cases an operator naively contributes to a vertex, yet that 
contribution does not lead to observable effects.
In such cases the \bf `X' \rm is replaced by an \bf `O' \rm.} 
\label{table-linear}
\end{center}
\end{table}


\section{The non-linear realization and operators of the 
electroweak chiral Lagrangian.}\label{sec-nonlinear}

It is possible that there is no physical Higgs boson, and the mechanism 
of spontaneous symmetry breaking is not part of the SM, but a part of its 
extension.  In such a scenario the full Lagrangian may be written as
\begin{equation}
\label{fulllagrangiannl}
{\cal L}_{\rm eff} = {\cal L}_{\rm SM} + \sum_i {\cal L}_i + \cdots \;.
\end{equation}
In contrast to the linearly realized Lagrangian (\ref{fulllagrangian}), the 
leading correction terms in the chiral Lagrangian are not suppressed by 
inverse powers of some high scale.

In order to proceed it is convenient to introduce some additional notation.
The charge-conjugate Higgs field may be written 
$\Phi^c = i\tau^2\Phi^\ast$.  Then
\begin{subequations}
\begin{eqnarray}
U & \equiv & \frac{\sqrt{2}}{v} \Big( \Phi^c , \Phi \Big)
\longrightarrow \id \;, \\
\dcov_\mu U & = & \partial_\mu U + i g T^a W_\mu^a U - i g^\prime U T^3 B_\mu 
\longrightarrow i g T^a W_\mu^a - i g^\prime T^3 B_\mu  \;, \\
T & \equiv & 2 U T^3 U^\dagger
\longrightarrow 2 T^3 \;, \\
V_\mu & \equiv & ( \dcov_\mu U ) U^\dagger
\longrightarrow \dcov_\mu U\;.
\end{eqnarray}
\end{subequations}
The arrows indicate the unitary-gauge expression.
In the notation of Appelquist and Wu\cite{AW} we present a list of chiral 
operators 
through energy-dimension four which conserve CP.  There are 
twelve such operators given by
\begin{subequations}
\label{twelve-chiral}
\begin{eqnarray}
\label{l1p}
{\cal L}_1^\prime 
  & = & \frac{\beta_1 v^2}{4} \Big[ \tr ( T V_\mu ) \Big]^2 \;,\\
\label{l1}
{\cal L}_1 & = &  \frac{\alpha_1 g g^\prime}{2} B_{\mu\nu} 
	\tr \Big( T W^{\mu\nu} \Big) \;,\\
\label{l2}
{\cal L}_2 & = & \frac{i \alpha_2 g^\prime}{2} B_{\mu\nu}
	\tr \Big( T [ V^\mu , V^\nu ] \Big) \;,\\
\label{l3}
{\cal L}_3 & = & i \alpha_3 g\,\tr \Big( W_{\mu\nu} [ V^\mu , V^\nu ] \Big) \;,\\
\label{l4}
{\cal L}_4 & = & \alpha_4 \Big[ \tr ( V_\mu V_\nu ) \Big]^2 \;,\\
\label{l5}
{\cal L}_5 & = & \alpha_5 \Big[ \tr ( V_\mu  V^\mu ) \Big]^2 \;,\\
\label{l6}
{\cal L}_6 & = & \alpha_6 \tr \Big( V_\mu V_\nu \Big) 
	\tr \Big( T V^\mu \Big) \tr \Big( T V^\nu \Big) \;,\\
\label{l7}
{\cal L}_7 & = & \alpha_7 \tr \Big( V_\mu V^\mu \Big) 
	\tr \Big( T V_\nu \Big) \tr \Big( T V^\nu \Big) \;,\\
\label{l8}
{\cal L}_8 & = & \frac{\alpha_8 g^2}{4} \Big[ \tr ( T W_{\mu\nu} ) \Big]^2 \;,\\
\label{l9}
{\cal L}_9 & = & \frac{i \alpha_9 g}{2} \tr \Big( T W_{\mu\nu} \Big) 
	\tr \Big( T [V^\mu , V^\nu ] \Big) \;,\\
\label{l10}
{\cal L}_{10} & = & \frac{\alpha_{10}}{2} 
	\Big[ \tr ( T V_\mu ) \tr ( T V_\nu ) \Big]^2 \;,\\
\label{l11}
{\cal L}_{11} & = & \alpha_{11}\, g\, \epsilon^{\mu\nu\rho\sigma}
\tr \Big( T V_\mu \Big) \tr \Big( V_\nu W_{\rho\sigma} \Big) \;.
\end{eqnarray}
\end{subequations}
The last operator, ${\cal L}_{11}$, violates parity, P.  The operators are
summarised in Table~\ref{table-chiral}, which employs the same format as 
Table~\ref{table-linear}.  Additionally Table~\ref{table-chiral} pairs each 
chiral operator with its linearly realized counterpart, four of which appear
in Sec.~\ref{sec-linear}.  The remainder, which occur at the 
energy-dimension-eight, -ten and -twelve level may be found elsewhere\cite{HHIS}.
\begin{table}[htbp]
\begin{center}
\begin{tabular}{|l||c||c|c|c|c|c|c|c|c|c|c|c|}
\hline
\raisebox{5mm}{${\cal L}_{\rm chiral}$} &
\makebox[.2cm]{}\raisebox{5mm}{${\cal O}^{(n)}_{\rm linear}$} & 
\rotate[l]{\makebox[1.8cm][c]{WW}} & \rotate[l]{\makebox[1.8cm][c]{ZZ}} & 
\rotate[l]{\makebox[1.8cm][c]{AZ}} & \rotate[l]{\makebox[1.8cm][c]{AA}} & 
\rotate[l]{\makebox[1.8cm][c]{WWZ}} & \rotate[l]{\makebox[1.8cm][c]{WWA}} & 
\rotate[l]{\makebox[1.8cm][c]{WWWW}} & \rotate[l]{\makebox[1.8cm][c]{WWZZ}} & 
\rotate[l]{\makebox[1.8cm][c]{WWZA}} & \rotate[l]{\makebox[1.8cm][c]{WWAA}} & 
\rotate[l]{\makebox[1.8cm][c]{ZZZZ}} \\ \hline \hline
\raisebox{-0.2cm}{${\cal L}_1^\prime $} &
\raisebox{-0.2cm}{$-\frac{4 \beta_1}{v^2} {\cal O}_{\Phi,1}$} 
&\makebox[0.4cm]{}&\cm&&&&&&&&\makebox[0.4cm]{}& \\[0.4cm] \hline
\raisebox{-0.2cm}{${\cal L}_1$} &
\raisebox{-0.2cm}{$\frac{4 \alpha_1}{v^2} {\cal O}_{BW}$}&&\cm&\cm&\cm&\cm&\cm&&&&&\\[0.4cm] \hline
\raisebox{-0.2cm}{${\cal L}_2 $} &
\raisebox{-0.2cm}{$\frac{8 \alpha_2 }{v^2} {\cal O}_{B}$}&&&&&\cm&\cm&&&&&\\[0.4cm] \hline
\raisebox{-0.2cm}{${\cal L}_3 $} &
\raisebox{-0.2cm}{$\frac{8 \alpha_3}{v^2} {\cal O}_{W}$} & 
&&&& \cm & \cm & \cm & \cm & \cm && \\[0.4cm] \hline
\raisebox{-0.2cm}{${\cal L}_4 $} &
\raisebox{-0.2cm}{$\frac{4 \alpha_4}{v^4} {\cal O}^{(8)}_{4}$} & 
&&&&&& \cm & \cm &&& \cm \\[0.4cm] \hline
\raisebox{-0.2cm}{${\cal L}_5 $} &
\raisebox{-0.2cm}{$\frac{16 \alpha_5}{v^4} {\cal O}^{(8)}_{5}$} & 
&&&&&& \cm & \cm &&& \cm \\[0.4cm] \hline
\raisebox{-0.2cm}{${\cal L}_6 $}& 
\raisebox{-0.2cm}{$-\frac{64 \alpha_6}{v^6} {\cal O}^{(10)}_{6}$} & 
&&&&&&& \cm &&& \cm \\[0.4cm] \hline
\raisebox{-0.2cm}{${\cal L}_7 $}&
\raisebox{-0.2cm}{$-\frac{64 \alpha_7}{v^6} {\cal O}^{(10)}_{7}$} & 
&&&&&&& \cm &&& \cm \\[0.4cm] \hline
\raisebox{-0.2cm}{${\cal L}_8 $} &
\raisebox{-0.2cm}{$-\frac{4 \alpha_8 } {v^4} {\cal O}^{(8)}_{8}$} & 
& \cm & \cm & \cm & \cm & \cm & \cm &&&& \\[0.4cm] \hline 
\raisebox{-0.2cm}{${\cal L}_9 $} &
\raisebox{-0.2cm}{$-\frac{16 \alpha_9 }{v^4} {\cal O}^{(8)}_{9}$} & 
&&&& \cm & \cm & \cm &&&& \\[0.4cm] \hline  
\raisebox{-0.2cm}{${\cal L}_{10} $} &
\raisebox{-0.2cm}{$\frac{128\, \alpha_{10} }{v^8} {\cal O}^{(12)}_{10}$} & 
&&&&&&&&&& \cm \\[0.4cm] \hline
\raisebox{-0.2cm}{${\cal L}_{11} $} &
\raisebox{-0.2cm}{$\frac{8 \alpha_{11}}{v^4} {\cal O}^{(8)}_{11}$} & 
&&&&\cm&&&& \cm && \\[0.4cm]
\hline
\end{tabular}
\vspace*{0.2cm}
\caption{Column one lists energy-dimension-four operators in the non-linear 
representation.  The linear-representation counterparts appear in the
second column.   For the definitions of the operators ${\cal O}^{(n)}_{i}$
the reader is referred to the text.
An \bf `X' \rm is used to indicate the the contribution of an individual 
operator to a particular vertex.}
\label{table-chiral}
\end{center}
\vspace{-.6cm}
\end{table}

Three of the chiral operators, ${\cal L}_{1}^\prime$, ${\cal L}_{1}$ and 
${\cal L}_{8}$, contribute to gauge-boson two-point-functions.  Like 
${\cal O}_{\Phi,1}$, ${\cal L}_{1}^\prime$ contributes only to the Z-mass term 
but not to the W-mass term and leads to a violation of the custodial symmetry.  
Through contributions to the charge form-factors $\baresq(q^2)$, $\barssq(q^2)$, 
$\bargzsq(q^2)$ and  $\bargsq(q^2)$ these three operators will contribute to 
the process \eeww.  None of the operators contributes to the WW 
two-point-function, hence, in contrast to the linear realization, no 
contribution will be made via the W-boson wave-function-renormalization factor,
and the t-channel terms are not modified.

In total six of the operators, ${\cal L}_{1}$, ${\cal L}_{2}$, ${\cal L}_{3}$,
${\cal L}_{8}$, ${\cal L}_{9}$ and ${\cal L}_{11}$,  contribute directly to 
three-gauge-boson vertices.  The non-zero corrections are
\begin{subequations}
\label{nonlinear-direct}
\begin{eqnarray}
\label{deltag1zdirectnon}
\Delta g_{1,{\rm direct}}^{Z, \rm nl} & = & \hatgzsq \alpha_3 \;, \\
\label{deltakappagammadirnon}
\Delta \kappa_{\gamma,{\rm direct}}^{\rm nl} & = & 
   \hatgsq \Big( - \alpha_1 + \alpha_2
  + \alpha_3 - \alpha_8 + \alpha_9 \Big)  \;, \\
\label{deltakappazdirnon}
\Delta \kappa_{Z,{\rm direct}}^{\rm nl} & = & 
   \hatgzsq \hatssq \Big( \alpha_1 - \alpha_2
   \Big) + \hatgsq \Big( \alpha_3 - \alpha_8 + \alpha_9 \Big)  \;, \\
\label{deltag5zdirectnon}
\Delta g_{5,{\rm direct}}^{Z, \rm nl} & = & \hatgzsq \alpha_{11} \;.
\end{eqnarray}
\end{subequations}
Notice that here, unlike Eqn.~(\ref{deltag1gammadirect}), 
$\Delta g_{1,{\rm direct}}^{\gamma, \rm nl}$ is trivially zero.  Furthermore 
notice that $\Delta \lambda_{\gamma,{\rm direct}}^{\rm nl}$ and 
$\Delta \lambda_{Z,{\rm direct}}^{\rm nl}$ are identically zero; these 
couplings are associated with energy-dimension-six operators, which are 
higher order corrections in the present scheme.  

 
\section{Four-fermion processes}\label{sec-eeff}

First we present the contributions of the linearly realized operators of
Table~\ref{table-linear} to the oblique parameters.  Due to the 
presence of $(q^2)^2$ terms in the gauge-boson two-point-functions the original 
derivative-based
definitions of $S$, $T$ and $U$ \cite{STU} are not convenient.  A more 
appropriate scheme for present purposes is where
the derivative is replaced by a finite difference\cite{HHKM} of the form
\begin{eqnarray}
\overline{\Pi}^{AB}_{T,V}(q^2) = {\overline{\Pi}^{AB}_T(q^2) 
- \overline{\Pi}^{AB}_T(m^2_V)\over q^2 -
m^2_V}\;.
\end{eqnarray}
Then
\begin{subequations}
\label{relationship}
\begin{eqnarray}\label{deltas}
\Delta S & \equiv & 16\pi\,{\cal R}\!e
\bigg[\Delta\overline{\Pi}^{3Q}_{T,\gamma}(m^2_Z) - \Delta\overline{\Pi}^{33}_{T,Z}(0)\bigg]
= - 4\pi \frac{v^2}{\Lambda^2}f_{BW}
\;,
\\ \label{deltat}
\Delta T & \equiv & \frac{4\sqrt{2} G_F}{\alpha} \,{\cal R}\!e
\bigg[\Delta\overline{\Pi}^{33}_{T}(0) - \Delta\overline{\Pi}^{11}_{T}(0)\bigg]
= - \frac{1}{2 \alpha} \frac{v^2}{\Lambda^2}f_{\Phi,1}
\;,
\\ \label{deltau}
\Delta U & \equiv & \makebox[0.12cm]{}
 16\pi\,{\cal R}\!e
\bigg[\Delta\overline{\Pi}^{33}_{T,Z}(0) - \Delta\overline{\Pi}^{11}_{T,W}(0)\bigg]
\makebox[0.13cm]{}
= 32\pi \frac{m_Z^2-m_W^2}{\Lambda^2}f_{DW}
\;,
\end{eqnarray}
\end{subequations}
where $S=S_{\rm SM} + \Delta S$, $T=T_{\rm SM} + \Delta T$ and
$U=U_{\rm SM} + \Delta U$.  Because the $f_{\Phi,1}$ and $f_{\Phi,4}$ 
contributions to the two-point functions are independent of $q^2$ they 
may contribute only to $T$.  The $f_{\Phi,4}$ contributions exactly cancel,
as was `predicted' during the discussion of Table~\ref{table-linear}.  
The $(q^2)^2$ terms in the two-point functions also lead to a non-standard 
running of the SM charge form-factors.  
The combination of $S$, $T$ and $U$ with the non-standard running 
leads to the convenient expressions
\begin{subequations}
\label{corrections}
\begin{eqnarray}\label{crctn_alpha}
\Delta \overline{\alpha}(q^2) & = & 
      - 8 \pi \hat{\alpha}^2 
	\frac{q^2}{\Lambda^2}\Big( f_{DW} + f_{DB} \Big)\;,
\\ \label{crctn_gzbar2}
\Delta \overline{g}_Z^2(q^2) & = & 
  	- 2 \hat{g}_Z^4 \frac{q^2}{\Lambda^2} \Big(\hat{c}^4 f_{DW} 
	+ \hat{s}^4 f_{DB} \Big) 
 	- \frac{1}{2} \hat{g}_Z^2 \frac{v^2}{\Lambda^2} f_{\Phi,1} \;,
\\ \nonumber
\Delta \overline{s}^2(q^2) & = & 
	 \frac{-\hat{s}^2\hat{c}^2}{\hat{c}^2-\hat{s}^2}\Bigg[ 
	8\pi\hat{\alpha}\frac{m_Z^2}{\Lambda^2}\Big( f_{DW} + f_{DB} \Big)
	+ \frac{m_Z^2}{\Lambda^2} f_{BW} 
	- \frac{1}{2}\frac{v^2}{\Lambda^2}f_{\Phi,1}
	\Bigg]
\\
\label{crctn_sbar2}
   	&& \makebox[3cm]{} + 8\pi\hat{\alpha}\frac{q^2-m_Z^2}{\Lambda^2} 
   	  \Big( \hat{c}^2 f_{DW} - \hat{s}^2 f_{DB}\Big)\;,
\\ \label{crctn_gwbar2}
\Delta \overline{g}_W^2(q^2) & = & 
	-8\pi\hat{\alpha}\hat{g}^2 \frac{m_Z^2}{\Lambda^2}f_{DB}
	- \hat{g}^2 \frac{ \Delta \overline {s}^2(m_Z^2)}{\hat{s}^2} 
	- \frac{1}{4} \hat{g}^4 \frac{v^2}{\Lambda^2}f_{BW} 
	- 2\hat{g}^4 \frac{q^2}{\Lambda^2}f_{DW} \;.
\end{eqnarray}
\end{subequations}
The `hatted' couplings satisfy the tree-level relationships
$\hat{e} \equiv \hat{g}\hat{s} \equiv \hat{g}_Z\hat{s}\hat{c}$ and 
$\hat{e}^2 \equiv 4\pi\hat{\alpha} $.  

The calculations may be repeated for the non-linear representation.  In this 
case the dependence on $q^2$ is at most linear, hence there is no non-standard 
running of the charge form-factors.  The oblique parameters are given by
\begin{subequations}
\label{relationship_nl}
\begin{eqnarray}
\label{deltas_nl}
\Delta S & = & -16\pi\alpha_1 \;, \\ 
\label{deltat_nl}
\Delta T & = & \frac{2}{\alpha}\beta_1\;, \\ 
\label{deltau_nl}
\Delta U & = & -16\pi\alpha_8 \;,
\end{eqnarray}
\end{subequations}
which agrees  with \cite{AW}.  The contributions to the charge form-factors 
may be written
\begin{subequations}
\label{corrections_nl}
\begin{eqnarray}
\label{crctn_alpha_nl}
\Delta \overline{\alpha}(q^2) & = & 0 \;, \\
\label{crctn_gzbar2_nl}
\Delta \overline{g}_Z^2(q^2) & = & 2 \hatgzsq \beta_1 \;, \\
\label{crctn_sbar2_nl}
\Delta \overline{s}^2(q^2) & = & -\frac{\hatcsq\hatssq}{\hatcsq-\hatssq}
\Big( 2\beta_1 + \hatgzsq \alpha_1 \Big) \;, \\
\label{crctn_gwbar2_nl}
\Delta \overline{g}_W^2(q^2) & = & -\hatgsq 
\frac{\Delta \overline{s}^2(m_Z^2)}{\hatssq} 
- \hat{g}^4 \Big( \alpha_1 + \alpha_8 \Big)  \;.
\end{eqnarray}
\end{subequations}

The low-energy data severely constrain those parameters of the chiral Lagrangian 
which contribute to $S$, $T$ and $U$; these constraints may be improved via 
improved measurements at low energies.  The corresponding parameters of the 
linearly realized Lagrangian are less stringently constrained at present.  
However, because some of their contributions to observables are enhanced by 
higher powers of $q^2$ these constraints may be expected to improve significantly
at LEP2 and beyond\cite{HMS}.


\section{Corrections to \eeww}\label{sec-eeww}

\begin{figure}[htb]
\epsfxsize = 4cm
\centerline{\epsffile{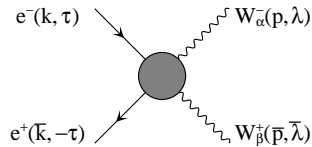}}
\vspace*{0.2cm}
\caption{The process $ee \rightarrow W^+W^-$\/ with momentum and helicity 
assignments.  All momenta are defined flowing to the right.  In the 
massless-electron limit $\overline{\tau}=-\tau$.}
\label{fig-eeww-blob} 
\end{figure}

The most general amplitude for \eeww\/, depicted in Figure~\ref{fig-eeww-blob}  
may be written
\begin{equation}
\label{general-form}
{\cal M}(k,\bar{k},\tau;p,\bar{p},\lambda,\bar{\lambda})
= \sum_{i=1}^{9} F_{i,\tau}(s,t)\, j_\mu(k,\bar{k},\tau) T_i^{\mu\alpha\beta} 
\epsilon^\ast_\alpha(p,\lambda) \epsilon^\ast_\beta(\bar{p},\bar{\lambda}) \;,
\end{equation}
where all dynamical information is contained in the scalar form-factors 
$F_{i,\tau}(s,t)$.  
The other factors in Eqn.~(\ref{general-form}) are of a purely 
kinematical nature; $\epsilon^\ast_\alpha(p,\lambda)$ and 
$\epsilon^\ast_\beta(\bar{p},\bar{\lambda})$ are the polarization vectors for 
the $W^-$ and $W^+$ bosons respectively, and $j_\mu(k,\bar{k},\tau)$
is the fermion current for massless electrons.  It is necessary to include 
nine tensors, $T_i^{\mu\alpha\beta}$, to be completely general. One possible 
choice is 
\begin{subequations}
\begin{eqnarray}
T_1^{\mu\alpha\beta} & = &  P^\mu g^{\alpha\beta} \;,\\
T_2^{\mu\alpha\beta} & = & \frac{-1}{m_W^2} P^\mu q^\alpha q^\beta \;,\\
T_3^{\mu\alpha\beta} & = & q^\alpha g^{\mu\beta} - q^\beta g^{\alpha\mu} \;,\\
T_4^{\mu\alpha\beta} & = & 
               i \Big( q^\alpha g^{\mu\beta} + q^\beta g^{\alpha\mu} \Big) \;,\\
T_5^{\mu\alpha\beta} & = & i \epsilon^{\mu\alpha\beta\rho} P_\rho \;,\\
T_6^{\mu\alpha\beta} & = & - \epsilon^{\mu\alpha\beta\rho} q_\rho \;,\\
T_7^{\mu\alpha\beta} & = & \frac{-1}{m_W^2} P^\mu 
                    \epsilon^{\alpha\beta\rho\sigma} q_\rho P_\sigma \;,\\
T_8^{\mu\alpha\beta} & = &  K^\beta g^{\alpha\mu} + K^\alpha g^{\mu\beta}\;,\\
T_9^{\mu\alpha\beta} & = &  \frac{i}{m_W^2} 
                            \Big( K^\alpha \epsilon^{\beta\mu\rho\sigma} 
             + K^\beta \epsilon^{\alpha\mu\rho\sigma} \Big) q_\rho P_\sigma\;,
\end{eqnarray}
\end{subequations}
where $P = p-\bar{p}$ and $K = k-\bar{k}$.  This completely determines the 
kinematics, hence we focus on the form factors $F_{i,\tau}(s,t)$.

We include only the tree-level contributions of the effective Lagrangians
from Sec.~\ref{sec-linear} and Sec.~\ref{sec-nonlinear}.
If we choose the renormalization conditions $\hatssq = \barssq(q^2)$ and 
$\hatesq = \baresq(q^2)$ the expressions for the $F_{i,\tau}(s,t)$ are 
somewhat simplified.  We write
\begin{equation}
F_{i,\tau}(s,t) = \frac{1}{s}Q\hatesq f_{i,\tau}^{\gamma}
	+ \frac{1}{s-m_Z^2 + is\frac{\Gamma_Z}{m_Z}\Theta(s)}
	 \big(I_3-\hatssq Q\big)\hatcsq\hatgzsq f_{i,\tau}^{Z}
	+\frac{1}{2t} I_3 \hatgsq f_{i,\tau}^{t}\;.
\label{big-simple}
\end{equation}
The amplitudes for \eeww\/ are completely determined once the lower-case 
form-factors, $f_i^X$, are specified.  The results for the SM, which are 
particularly simple, may be found in Table~\ref{table-smff}.
\begin{table}[htb]
\begin{center}
\begin{tabular}{|c||c|c|c|c|c|c|c|c|c|}
\hline
&$i=1$ &$i=2$ &$i=3$ &$i=4$ &$i=5$ &$i=6$ &$i=7$ &$i=8$ &$i=9$ \\[0.1cm] 
\hline \hline 
$f_{i,\pm}^{\gamma}$ & 1 & 0 & 2 & 0 & 0 & 0 & 0 & 0 & 0 \\[0.1cm] \hline
$f_{i,\pm}^{Z}$      & 1 & 0 & 2 & 0 & 0 & 0 & 0 & 0 & 0 \\[0.1cm] \hline
$f_{i, -}^{t}$        & 1 & 0 & 2 & 0 & 1 & 0 & 0 & 1 & 0 \\[0.1cm] \hline
\end{tabular}
\vspace{0.2cm}
\caption{Explicit values for the form factors of the SM at the tree level.}
\label{table-smff}
\end{center}
\end{table}
These values guarantee the gauge cancellations which prevent unitarity 
violations in amplitudes which involve one or more longitudinally polarized 
W boson.

We may calculate the form-factors with effective-Lagrangian contributions.  In
general we expect results that will spoil the gauge cancellations of the SM.
The non-zero form-factors in the linear representation, for the photonic 
contributions, may be written
\begin{subequations}
\label{ff-gamma-linear}
\begin{eqnarray}
f_{1,\pm}^{\gamma} & = & 1 +\frac{\Delta\baresq(s)}{\hatesq} +
   \hatgsq\frac{s}{\Lambda^2} \Big(-f_{DW}+\frac{3}{4}f_{WWW} \Big) \;,
\\
f_{2,\pm}^{\gamma} & = &\frac{3}{2}\hatgsq\frac{m_W^2}{\Lambda^2}
	 \Big(-4f_{DW} +f_{WWW} \Big) \;,
\\
\nonumber
f_{3,\pm}^{\gamma} & = & 2 + 2\frac{\Delta\baresq(s)}{\hatesq} +
   2\hatgsq \frac{2s-3m_W^2}{\Lambda^2}f_{DW}
	+\frac{3}{2}\hatgsq\frac{m_W^2}{\Lambda^2} f_{WWW}
\\ && \mbox{}
  +\frac{1}{2} \frac{m_W^2}{\Lambda^2}\Big( f_W -2 f_{BW} + f_B\Big)   \;,
\end{eqnarray}
\end{subequations}
where $\Delta\baresq(q^2)$ is obtained from (\ref{crctn_alpha}).  For the Z-boson
\begin{subequations}
\label{ff-z-linear}
\begin{eqnarray}
f_{1,\pm}^{Z} & = & 1 +\frac{\Delta\bargzsq(s)}{\hatgzsq} +
  \hatgsq \frac{s}{\Lambda^2}\Big( -f_{DW} + \frac{3}{4} f_{WWW}\Big) 
  + \frac{1}{2}\frac{m_Z^2}{\Lambda^2}f_W\;,
\\
f_{2,\pm}^{Z} & = & \frac{3}{2}\hatgsq\frac{m_W^2}{\Lambda^2}
	\Big(-4f_{DW}+f_{WWW}\Big)  \;,
\\
\nonumber
f_{3,\pm}^{Z} & = & 2 + 2\frac{\Delta\bargzsq(s)}{\hatgzsq} +
2\hatgsq \frac{2s-3m_W^2}{\Lambda^2}f_{DW}
	+\frac{3}{2}\hatgsq\frac{m_W^2}{\Lambda^2} f_{WWW}
\\ && \makebox[0cm]{}
	+ \frac{1}{2}\frac{m_Z^2}{\Lambda^2}\bigg[\Big(1+\hatcsq\Big)
	f_W +2\hatssq f_{BW} - \hatssq f_B \bigg]    \;,  
\end{eqnarray}
\end{subequations}
with $\Delta\bargzsq(q^2)$ given by (\ref{crctn_gzbar2}).  In the t-channel
\label{ff-t-linear}
\begin{equation}
f_{1,-}^{t} \frac{1}{2} = f_{3,-}^{t\,({\rm eff})} = 
f_{5,-}^{t} = f_{8,-}^{t\,({\rm eff})} 
  = \frac{\bargwsq(m_W^2)}{\hatgsq}\;,
\end{equation}
with $\bargwsq(m_W^2)$ given by Eqn.~(\ref{crctn_gwbar2}).

The calculation may be repeated for the chiral Lagrangian of 
Sec.~\ref{sec-nonlinear}.  For the $f_{i,\tau}^{\gamma}$ we 
find
\begin{subequations}
\begin{eqnarray}
\label{ff-gamma-nonlinear}
f_{1,\pm}^{\gamma} & = & 1 \;,\\
f_{3,\pm}^{\gamma} & = & 2 + 
   \hatgsq ( -\alpha_1 + \alpha_2 + \alpha_3 - \alpha_8 + \alpha_9 )  \;.
\end{eqnarray}
\end{subequations}
while for $f_{i,\tau}^{Z}$
\begin{subequations}
\label{ff-z-nonlinear}
\begin{eqnarray}
f_{1,\pm}^{Z} & = & 1+2\beta_1+ \hatgzsq \alpha_3 \;,
\\
f_{3,\pm}^{Z} & = & 2 + 4\beta_1 
  + \hatgzsq\hatssq ( \alpha_1 - \alpha_2 ) + \hatgzsq(1+\hatcsq)\alpha_3
  + \hatgsq(-\alpha_8 + \alpha_9)\;,
\\
f_{5,\pm}^{Z} & = & \hatgzsq \alpha_{11}\;.
\end{eqnarray}
\end{subequations}
For the t-channel form-factors,
\label{ff-t-nonlinear}
\begin{eqnarray}
f_{1,-}^{t} = \frac{1}{2} f_{3,-}^{t\,({\rm eff})} =
f_{5,-}^{t} = f_{8,-}^{t\,({\rm eff})} 
 & = & 1 +  \frac{2 \hatgsq}{\hatcsq-\hatssq}\Big( \hatcsq \beta_1 
+ \hatesq \alpha_1 \Big) - \hat{g}^4 \alpha_8  \;,
\end{eqnarray}

The above contributions in general violate the gauge cancellations of the SM.
There is not sufficient space for a detailed discussion.  Notice, however, that
the results are consistent with electromagnetic gauge invariance.  We may 
calculate $g_1^\gamma = f_1^\gamma - (q^2/2m_W^2)f_2^\gamma$, and we anticipate
that the result should be $g_1^\gamma = 1$ for on-shell photons.  For the 
linearly realized scenario $\Delta g_1^\gamma = g_1^\gamma - 1$ is proportional
to $q^2$, while for the chiral Lagrangian $\Delta g_1^\gamma = 0$; both satisfy
the gauge-invariance condition.


\section{Conclusions}\label{sec-conclusions}

I have outlined the contributions of the effective theories to processes with 
four external fermions.  In this sector extensive numerical studies have been
completed which include complete SM radiative corrections\cite{HISZnew,HMS}.

I also presented complete amplitudes for \eeww.\/  While in either realization 
of the Higgs sector seven operators contribute to these amplitudes, in the 
linear (non-linear) realization four (three) are already constrained via 
four-fermion processes.  Hence the number of free parameters is very manageable, 
and this may prove to be an extremely useful tool for the analysis of LEP2 and 
linear collider data.  

It is expected that LEP2 will not be sensitive to the SM radiative corrections
to \eeww,\/ hence the present analysis, which treats the non-standard effects as
large compared to SM radiative corrections, is of immediate interest.  However,
eventually it will be necessary to include complete SM corrections.  The analysis
presented is sufficiently general to include the complete corrections; 
Eqn.~\ref{general-form} is completely general, though the
expressions for the $F_{i,\tau}(s,t)$ become more complicated than 
Eqn.~\ref{big-simple}.


\newpage
\section*{Acknowledgements}
\noindent 
I would like to thank my collaborators, Kaoru Hagiwara, Tsutomu Hatsukano and
Satoshi Ishihara.  I would also like to thank the organisers of YKIS95 for
the priveledge of speaking at a splendid conference.  



\begin{thebibliography}{99}
%
\bibitem{HISZnew}
K.~Hagiwara, S.~Ishihara, R.~Szalapski and D.~Zeppenfeld,
Phys. Rev. {\bf D48} (1993) 2182.
%
\bibitem{HMS}
K.~Hagiwara, S.~Matsumoto and R.~Szalapski, Phys. Lett. {\bf B357} (1995) 
411-418.
%
\bibitem{AW}
T.~Appelquist and G.H.~Wu, Phys. Rev. {\bf D48} (1993) 3235.
%
\bibitem{HHIS}
K.~Hagiwara, T.~Hatsukano, S.~Ishihara and R.~Szalapski, preprint in preparation.
%
\bibitem{STU}
M.E.~Peskin and T.~Takeuchi, Phys. Rev. Lett. {\bf 65} (1990) 964; 
 Phys. Rev. {\bf D46} (1992) 381.
%
\bibitem{HHKM}
K.~Hagiwara, D.~Haidt, C.~S.~Kim and S.~Matsumoto, 
Z.~Phys.~{\bf C64} (1994) 559.
%
\bibitem{HPZH}
K.~Hagiwara, R.D.~Peccei, D.~Zeppenfeld and K.~Hikasa, Nucl. Phys. {\bf B282}
(1987) 253.
%
\end{thebibliography}
\end{document}